\documentclass{aastex}
\usepackage{spr-astr-addons}

\RequirePackage{color}

\begin{document}

\title{Patterns of galaxy spin directions in SDSS and Pan-STARRS show parity violation and multipoles}
\shorttitle{Spin direction asymmetry}
\shortauthors{Lior Shamir}

\author{Lior Shamir\altaffilmark{1}}
\affil{Kansas State University, Manhattan, KS 66502 \\ \small lshamir@mtu.edu}

\begin{abstract}
The distribution of spin directions of $\sim6.4\cdot10^4$ SDSS spiral galaxies with spectra was examined, and compared to the distribution of  $\sim3.3\cdot10^4$ Pan-STARRS galaxies. The analysis shows a statistically significant asymmetry between the number of SDSS galaxies with opposite spin directions, and the magnitude and direction of the asymmetry changes with the direction of observation and with the redshift. The redshift dependence shows that the distribution of the spin direction of SDSS galaxies becomes more asymmetric as the redshift gets higher. Fitting the distribution of the galaxy spin directions to a quadrupole alignment provides fitness with statistical significance $>5\sigma$, which grows to $>8\sigma$ when just galaxies with $z>0.15$ are used. Similar analysis with Pan-STARRS galaxies provides dipole and quadrupole alignments nearly identical to the analysis of SDSS galaxies, showing that the source of the asymmetry is not necessarily a certain unknown flaw in a specific telescope system. While these observations are clearly provocative, there is no known error that could exhibit itself in such form. The data analysis process is fully automatic, and uses deterministic and symmetric algorithms with defined rules. It does not involve either manual analysis that can lead to human perceptual bias, or machine learning that can capture human biases or other subtle differences that are difficult to identify due to the complex nature of machine learning processes. Also, an error in the galaxy annotation process is expected to show consistent bias in all parts of the sky, rather than change with the direction of observation to form a clear and definable pattern. 
\end{abstract}

\maketitle

\section{Introduction}
\label{introduction}

Spiral galaxies are unique astronomical objects in the sense that their visual appearance depends on the perspective of the observer. Since the spin patterns of spiral galaxies (clockwise or counterclockwise) are expected to be randomly distributed, in a sufficiently large universe no difference between clockwise and counterclockwise galaxies is expected. However, analysis of large datasets of spiral galaxies showed photometric differences between spiral galaxies with clockwise spin patterns and spiral galaxies with counterclockwise spin patterns \citep{shamir2013color,shamir2016asymmetry,shamir2017colour,shamir2017photometric,shamir2017large,shamir2020asymmetry}. Early attempts to identify differences between clockwise and counterclockwise galaxies did not identify statistically significant asymmetry \citep{iye1991catalog,land2008galaxy}. However, these experiments were based on much smaller datasets of just a few thousand galaxies \citep{iye1991catalog}, or on heavily biased manual classification performed by untrained volunteers \citep{land2008galaxy}. Experiments using manually annotated galaxies \citep{longo2011detection} and automatically annotated galaxies \citep{shamir2012handedness,shamir2020large,shamir2020distribution} showed statistically significant differences between the number of clockwise and counterclockwise spiral galaxies in Sloan Digital Sky Survey (SDSS).

First evidence of photometric differences between clockwise and counterclockwise galaxies were observed using SDSS galaxies \citep{shamir2013color}. Machine learning algorithms showed accuracy much higher than mere chance in identifying the spin pattern of the galaxy by its photometric variables, showing a statistically significant link between the photometry of the galaxy and its spin pattern \citep{shamir2016asymmetry}. That was shown with datasets of both manually and automatically annotated galaxies \citep{shamir2016asymmetry}. The same datasets also showed that the the difference between the number of clockwise and counterclockwise galaxies changed with the redshift \citep{shamir2016v1,shamir2020distribution}.

Experiments with a much larger dataset of $\sim1.62\times10^5$ automatically annotated SDSS galaxies \citep{kuminski2016computer} showed statistically significant photometric differences between clockwise and counterclockwise galaxies. The experiments showed color \citep{shamir2017colour} and magnitude \citep{shamir2017photometric} differences that change with the direction of observation, and have a cosine dependence with the RA \citep{shamir2017colour,shamir2017photometric}. With SDSS data, maximum asymmetry was observed in the RA range of $(120^o,210^o)$.

Analysis of SDSS and PanSTARRS galaxies showed that data collected by both telescopes exhibit the same asymmetry, as well as the same asymmetry pattern \citep{shamir2017large}. In both telescopes, the asymmetry changed with the direction of observation. Analysis of the COSMOS field also showed asymmetry that agreed with the asymmetry observed by Pan-STARRS and SDSS \citep{shamir2020asymmetry}. While COSMOS is a deeper field, it is also much smaller than SDSS and Pan-STARRS, and contains less galaxies, therefore leading to weaker statistical signal of the asymmetry. A third dataset of $\sim$40K manually classified SDSS galaxies also showed the same pattern \citep{shamir2017large}. The observation that the asymmetry was identified by three different telescopes can be viewed as a certain indication that no error in the photometric pipeline of a specific sky survey is the cause of the asymmetry. A software error is also unlikely, as such error would have exhibited itself in the form of consistent bias in the entire sky, while the actual observations showed a link between the magnitude (and direction) of the asymmetry and the direction of observation. Additionally, manually annotated galaxies showed results that were in excellent agreement with the automatically annotated data \citep{shamir2017large}.

Smaller-scale observations showed that neighboring galaxies tend to have orthogonal spin patterns, which can be explained by gravitational interactions leading to faster merging of systems of galaxies with the same spin direction \citep{sofue1992spins,puerari1997relative}. Other observations provide evidence of correlation between neighboring galaxies and their spin patterns \citep{slosar2009galaxy}, even in cases of galaxies that are too far from each other to have gravitational interactions \citep{lee2019mysterious}. These observations show certain evidence of patterns at the cosmological scale reflected by the distribution of galaxy spin directions \citep{lee2019mysterious}. 

While it is difficult to explain these observations with the standard cosmological models, theoretical explanations that do not mandate the violation of the basic cosmological assumptions have been proposed \citep{yu2020probing,biagetti2020primordial}. Here I analyze $\sim6.4\cdot10^4$ SDSS galaxies with spectra and  $\sim3.3\cdot10^4$  Pan-STARRS galaxies to identify changes in the population of clockwise and counterclockwise galaxies based on their position in the sky and their redshift, as well as to compare the asymmetry observed by two independent digital sky surveys.

\section{Data}
\label{data}

The initial dataset is galaxies with spectra from the SDSS DR14. All objects with class ``GALAXY'' in the SpecObjAll table were selected in the initial query, providing a set of 2,644,145 objects with spectra identified as galaxies by SDSS pipeline. Since many of the objects are too small to allow any morphological analysis, a subset of these galaxies was selected such that the Petrosian radius computed on the g band was greater than 5.5''. That selection reduced the number of galaxies with spectra to a subset of 589,049 galaxies. The distribution of the galaxies in different RA and redshift ranges is specified in Table~\ref{distribution_table}.

\begin{table}
{
\scriptsize
\begin{tabular}{|l|c|c|c|c|}
\hline
z &  0$^o$-120$^o$    &  120$^o$-240$^o$  & 240$^o$-360$^o$ & Total \\      
\hline
0-0.05   &    9,655   &   64,092  &  10,862     &    84,609    \\
0.05-0.1 &   14,746  &  104,339   & 21,098   &   140,183    \\
0.1-0.15 &    12,142  &  67,712  &  13,797   &    93,651    \\   
0.15-0.2  &  7,757    & 33,957  &  8,360     &     50,074    \\
$>$0.2  &  47,456    & 134,706    & 36,473    &   218,635    \\
\hline
Total      &  91,965 &  406,185  & 90,899   & 589,049  \\
\hline        
\end{tabular}
\caption{Distribution of the galaxies in the dataset by redshift and RA ranges.}
\label{distribution_table}
}
\end{table}

The galaxy images were fetched from SDSS Skyserver using the ``cutout'' web service, and the output images were 120$\times$120 color JPEG images. To make sure that the entire galaxy fits inside the image, if more than 25\% of the pixels on the edge of the image were bright pixels (with grayscale value greater than 125), the image was downscaled and downloaded again until the number of bright pixels was less than 25\% of the total number of pixels on the edge. The initial scale of the image was 0.1 arcseconds per pixel, and it was reduced by 0.01 arcseconds per pixel in each iteration, until the galaxy fits in the frame \citep{kuminski2016computer}.

\begin{figure*}
\includegraphics[scale=1.0]{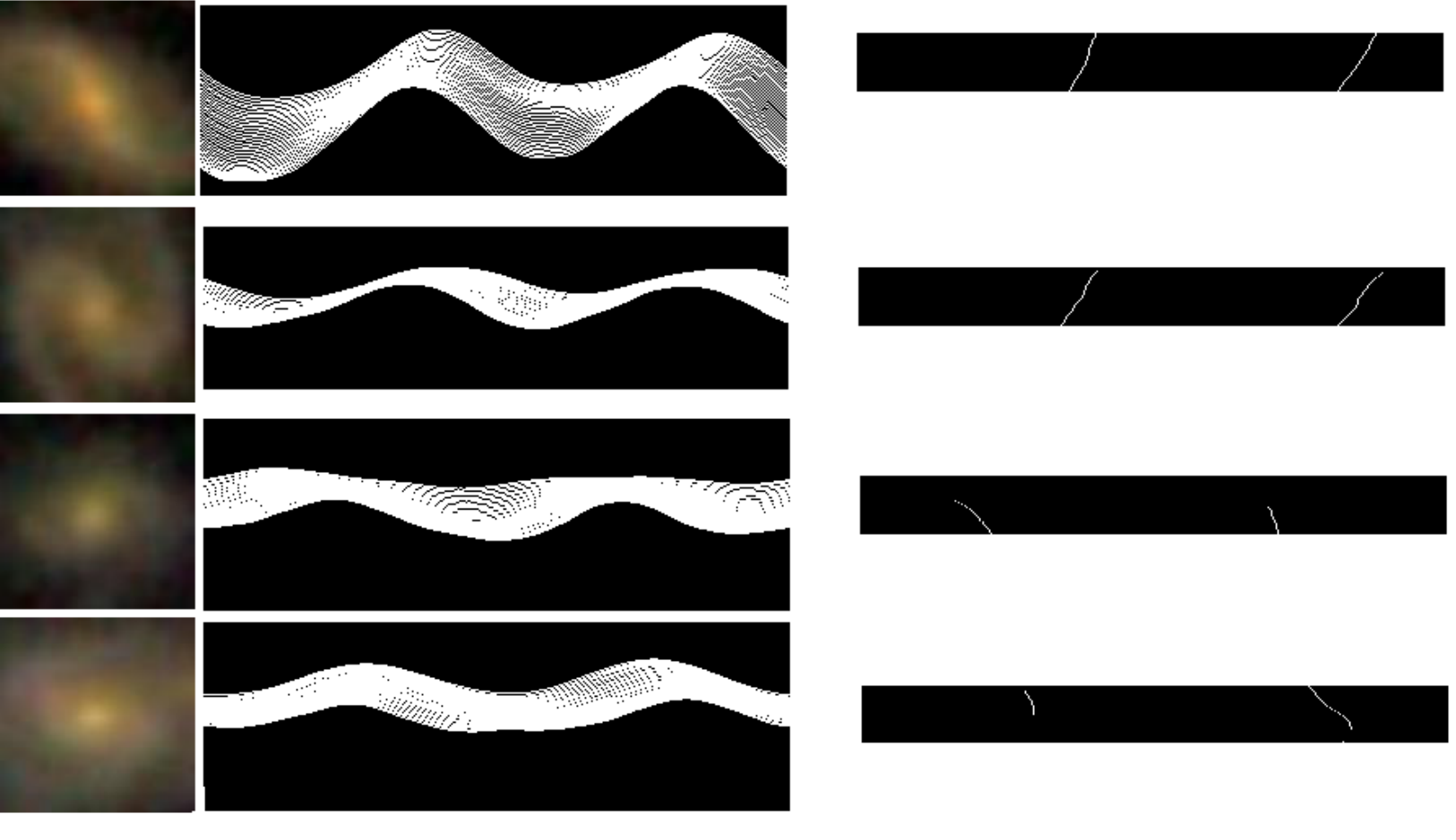}
\caption{Original images (left), their corresponding radial intensity plots (middle), and the peaks detected in the radial intensity plots (right). The sign of the lines of the peaks determines the curve of the arms of the galaxy, and consequently its direction of rotation}
\label{redial_intensity_plots}
\end{figure*}

The galaxies were classified into galaxies with clockwise spin patterns and galaxies with counterclockwise spin patterns by using the Ganalyzer tool \citep{shamir2011ganalyzer,ganalyzer_ascl}, as was done in \citep{shamir2012handedness,shamir2013color,hoehn2014characteristics,dojcsak2014quantitative,shamir2016asymmetry,shamir2017colour,shamir2017photometric,shamir2017large,shamir2020large}. Ganalyzer is a simple deterministic algorithm that is based on converting a galaxy image into its radial intensity plot \citep{shamir2011ganalyzer}, followed by peak detection in different radial distances from the center of the galaxy. The value of the pixel $(x,y)$ in the radial intensity plot is the median value of the 5$\times$5 pixels around $(O_x+\sin(\theta) \cdot r,O_y-\cos(\theta)\cdot r)$ in the original image, where $\theta$ is the polar angle and {\it r} is the radial distance. Because the arm of the galaxy is expected to be brighter than other parts of the galaxy at the same distance from the galaxy center, the groups of peaks in the radial intensity plot are expected to be associated with the arms of the galaxy. Figure~\ref{redial_intensity_plots} shows examples of original galaxy images, their corresponding radial intensity plots, and the peaks detected in the radial intensity plots.

The sign of the linear regression of the peaks determines the curve of the arms, and consequently the rotation direction of the galaxy. Obviously, not all galaxies are spiral, and not all spiral galaxies have identifiable spin patterns. To have good identification of the direction of the peaks, at least 30 peaks need to be detected in the radial intensity plot of the galaxy, otherwise the galaxy is labeled as unidentifiable and rejected from the analysis. The algorithm is described in detail in \citep{shamir2011ganalyzer}, as well as in \citep{shamir2012handedness,shamir2013color,hoehn2014characteristics,dojcsak2014quantitative,shamir2016asymmetry,shamir2017colour,shamir2017photometric}.

An important advantage of Ganalyzer is that it is a model-driven algorithm that is not based on machine learning. Therefore, human bias or other learning bias such as the part of the sky from which the training samples were taken cannot affect the performance or behavior of the algorithm. Ganalyzer is a straightforward deterministic algorithm that is not based on complex non-intuitive rules often used by machine learning algorithms, especially with deep neural networks. Ganalyzer is therefore not subjected to any training bias, as no training is required in any stage. The rules are designed in a fully symmetric manner, so that no bias of the algorithm to a certain type of spin pattern is allowed by the code.

\begin{figure}[h]
\includegraphics[scale=0.62]{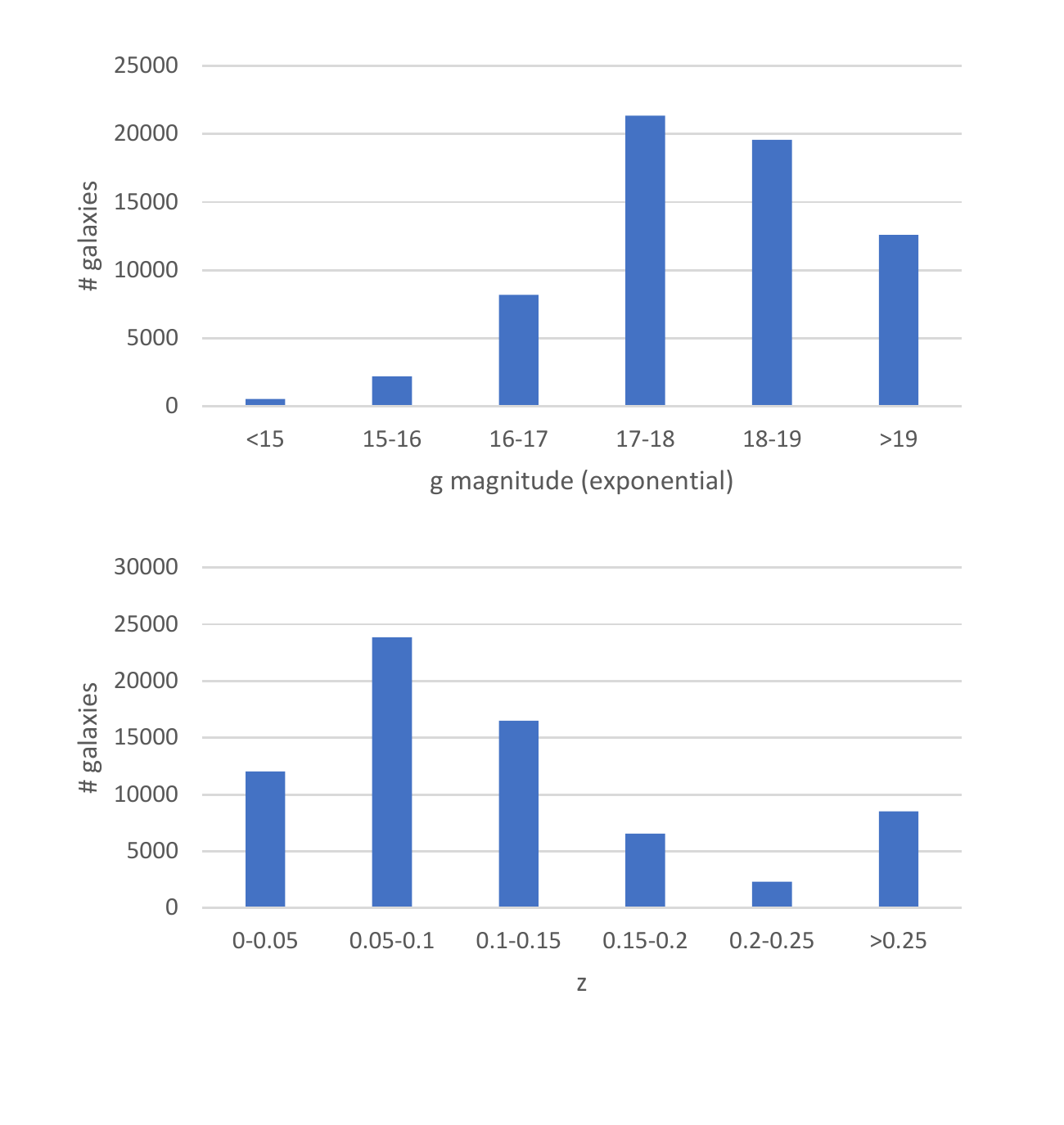}
\caption{The distribution of the exponential magnitude (g) and redshift of the galaxies used in the dataset.}
\label{distribution}
\end{figure}

The process resulted in a dataset of 32,055 galaxies with clockwise spin patterns, and 32,501 galaxies with counterclockwise spin patterns. Visual inspection of 100 randomly selected galaxies showed no cases of missclassified galaxies, indicating that the dataset is reasonably clean. SDSS can sometimes assign more than one object ID to the same object, especially in the case of extended objects. Removing these duplications provided a dataset of 63,693 galaxies, 31,666 clockwise galaxies and 32,027 are counterclockwise. The color and redshift distribution of the entire dataset population is shown in Figure~\ref{distribution}.


\begin{figure}[h]
\includegraphics[scale=0.60]{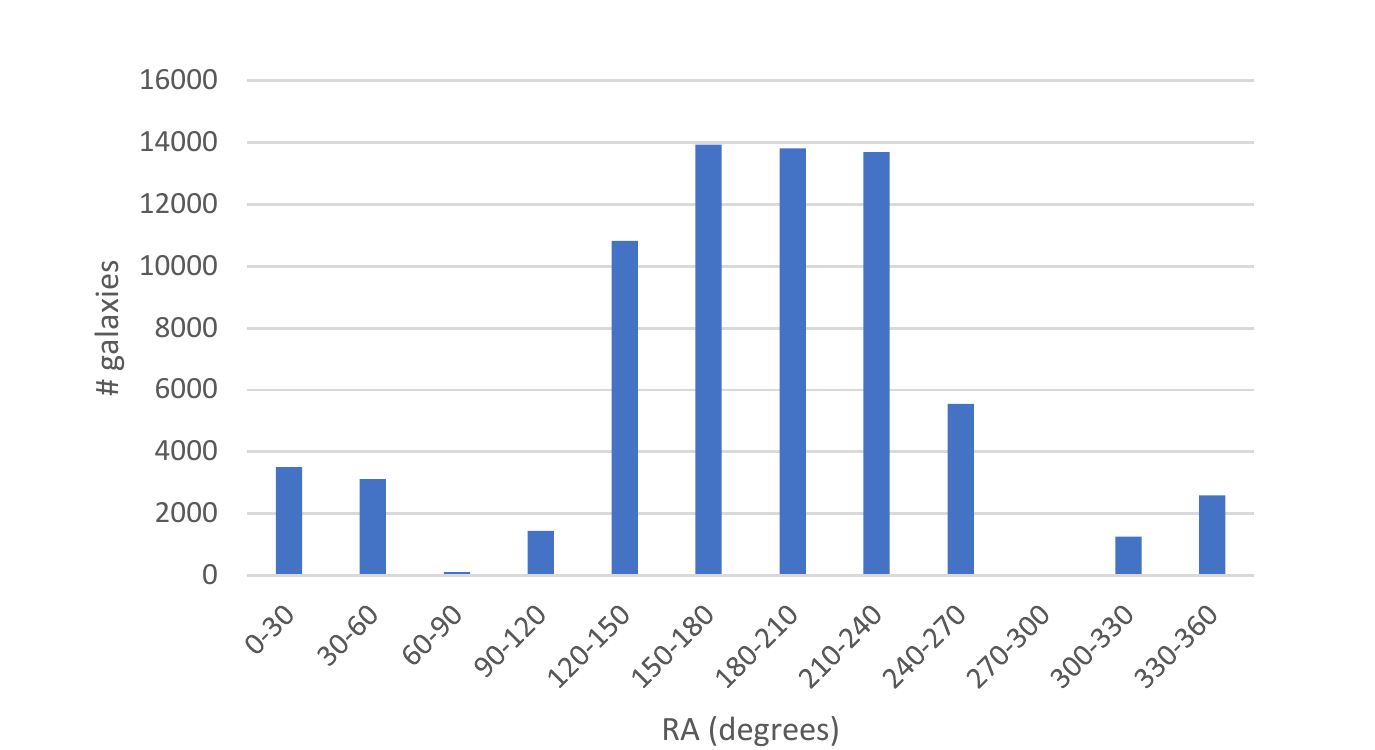}
\caption{The distribution of the galaxies in the different RA ranges.}
\label{distribution_ra}
\end{figure}

The galaxies with spectroscopy are not distributed uniformly in the sky covered by SDSS, and some parts of the sky contain more spectroscopic and photometric objects than others in the SDSS database. Figure~\ref{distribution_ra} shows the number of galaxies in each RA range. As the graph shows, the number of galaxies with spectra is particularly low in the RA ranges of $(60^o,90^o)$ and $(270^o,300^o)$.  

\section{Results}
\label{results}

Following the observations described in \citep{shamir2017large}, the galaxies were divided into four groups based on their right ascension. The division of the sky into regions is based on the observation that the asymmetry between clockwise and counterclockwise galaxies in SDSS and Pan-STARRS depends on the direction of observation \citep{shamir2017colour,shamir2017photometric,shamir2017large}, and the asymmetry is strongest in the RA range of (120$^o$,210$^o$). Mirroring the galaxy images and repeating the same analysis leads to the exact opposite results, which is expected since Ganalyzer is a symmetric and deterministic algorithm, and therefore the annotation of a certain galaxy image is the opposite of the annotation of its mirrored image. Table~\ref{directions} shows the number of clockwise and counterclockwise galaxies in each direction of observation.



\begin{table}
{
\scriptsize
\begin{tabular}{|l|c|c|c|c|c|}
\hline
RA &    cw    &  ccw  & ${cw}\over{cw+ccw}$ & P value & q-value \\      
\hline
120$^o$-210$^o$     & 17876 & 18391 & 0.493    & 0.0034  & 0.0136  \\
$>300^o<30^o$   & 3064 & 3219    & 0.487    & 0.026    & 0.104 \\
30$^o$-120$^o$      & 1665 & 1560     & 0.516      &	0.033  & 0.132 \\
210$^o$-300$^o$    & 9061 & 8857     &  0.506   & 0.065        & 0.258  \\
\hline
\end{tabular}
\caption{The distribution of the clockwise and counterclockwise galaxies in different RA ranges. The P value of the binomial distribution shows the probability that the asymmetry between the number of clockwise and counterclockwise galaxies occurs by chance. The q-value is the Bonferroni-corrected P value.}
\label{directions}
}
\end{table}

The table shows statistically significant difference between the number of clockwise galaxies and the number of counterclockwise galaxies in the RA range of $(120^o,210^o)$, the same RA range that showed photometric asymmetry between clockwise and counterclockwise galaxies \citep{shamir2017large}. The statistical power of the parity violation in that RA range remains significant also after applying the Bonferonni correction. The same RA range in the opposite hemisphere ($>300^o , <30^o$) also shows a higher population of counterclockwise galaxies, but when applying the Bonferonni correction the statistical significance drops below a discovery level. It should be noted that the RA range ($ > 300^o < 30^o$) has a much lower number of galaxies, which can have a major impact on the lower statistical significance. The other two RA ranges show higher numbers of clockwise galaxies, but the statistical significance of the difference is lower.  The analysis is obviously limited by ignoring the declination of the galaxies, but it shows that when the number of galaxies is sufficiently large, a difference in spin directions of galaxies in the same field can be identified.  

Table~\ref{dis_120_210} shows the distribution of clockwise and counterclockwise SDSS galaxies in the RA range of $(120^o,210^o)$ and different redshift ranges. The table shows the number of clockwise and counterclockwise galaxies in each redshift range. The P value shows the probability to have such distribution by chance, as determined by the accumulative binomial distribution assuming that the probability of a galaxy to have a clockwise or counterclockwise spin is exactly 0.5. As the table shows, the asymmetry between the number of clockwise and counterclockwise galaxies increases with the redshift. While the asymmetry between clockwise and counterclockwise galaxies is insignificant in redshift range of 0-0.05, the asymmetry grows consistently with the redshift range. That cannot be explained by changes in the performance of the galaxy classification algorithm, as lower performance of the algorithm should shift the difference closer to random distribution, which is 0.5, and in any case should have also affected the other RA ranges.

\begin{table}
{
\scriptsize
\begin{tabular}{|l|c|c|c|c|}
\hline
z &    cw    &  ccw  & ${cw}\over{cw+ccw}$ & P value \\      
\hline
0-0.05     &	3216 &	 3180 & 	0.5003	& 0.698  \\
0.05-0.1 &	6240	& 6270 &	0.498 &	0.4 \\
0.1-0.15 &	4236 &	4273 & 	0.496 &	0.285  \\
0.15-0.2 &	1586 &	1716 & 	0.479 &	0.008  \\
0.2 - 0.5 &	2598 &	 2952 &	 0.469 &	$1.07\cdot10^{-6}$  \\

\hline
Total & 17,876 & 18,391 & 0.493 & 0.0034 \\
\hline
\end{tabular}
\caption{The distribution of the clockwise and counterclockwise SDSS galaxies in different redshift ranges in the RA range of $(120^o,210^o)$. The P value of the accumulative binomial distribution shows the probability that the asymmetry between the number of clockwise and counterclockwise galaxies occurs by chance.}
\label{dis_120_210}
}
\end{table}

The Pearson correlation between each of the five redshift ranges and the asymmetry in each redshift range is $\sim$-0.951 (P$<$0.013). The Pearson correlation can also be computed without separating the dataset into redshift ranges. Assigning all clockwise galaxies to 1 and all counterclockwise galaxies to -1 provides a sequence of 63,693 pairs of values such that the redhisft of each galaxy is paired with 1 or -1, based on the spin direction of the galaxy. The Pearson correlation between these pairs of values is -0.01546. The correlation coefficient is low, as expected, but when the number of pairs in the dataset is 63,693 the probability of having such correlation by chance is (P$\simeq$0.00009). The correlation between the rotation direction (1 or -1) and the redshift in the RA range $(120^o,210^o)$ is -0.02329. With the 36,267 galaxies in that RA range the probability to get such correlation by chance is (P$<$0.0001). That shows that in the RA range $(120^o,210^o)$ the population of counterclockwise galaxies observed by SDSS increases with the redshift.

The link between the asymmetry and the redshift might be related to the size of the galaxies, as SDSS galaxies are generally expected to be smaller as their redshift gets higher. As mentioned in Section~\ref{data}, all galaxies used in the experiment have Petrosian radius of at least 5.5'', measured on the g band. That is expected to ensure that the galaxies are of sufficient size to analyze their morphology. To further test a possible link between the size of the galaxies and their redshift, galaxies with redshift $<0.15$ separated by their radius were used. As shown in Table~\ref{dis_120_210}, galaxies with redshift $<0.15$ in the RA range of $(120^o,210^o)$ are not expected to have significant asymmetry regardless of their redshift, and therefore a link between the asymmetry and their size inside that redshift range would be an indication that the increase in asymmetry shown in Table~\ref{dis_120_210} is driven by the size of the galaxies rather than their redshift. Table~\ref{by_radius} shows the number of the clockwise and counterclockwise galaxies in different Petrosian radii in the RA range of $(120^o,210^o)$ and $z<0.15$. As the table shows, the asymmetry does not change with the size of the galaxies, even for the smallest galaxies in the dataset with Petrosian radius in the range of 5.5'' to 6''. That shows that the size of the galaxy does not have a substantial impact on the asymmetry between the number of clockwise and counterclockwise galaxies in SDSS.

\begin{table}
{
\scriptsize
\begin{tabular}{|l|c|c|c|c|}
\hline
Radius ('') &    cw    &  ccw  & ${cw}\over{cw+ccw}$ \\      
\hline
$5.5-\infty$ & 13692   &   13723 & 0.499 \\
$5.5-8$  &  9398  &     9479    & 0.497 \\
$5.5-7$  & 7765    &   7776  & 0.499 \\
$5.5-6$  & 5454     &  5473  & 0.499 \\
\hline
\end{tabular}
\caption{The distribution of the clockwise and counterclockwise SDSS galaxies in the RA range of $(120^o,210^o)$ and the redshift range of $z<0.15$, and different Petrosian radii.}
\label{by_radius}
}
\end{table}

In addition to the RA range of $(120^o,210^o)$, the asymmetry in different redshift ranges was tested also for the other three RA ranges. The number of clockwise and counterclockwise SDSS galaxies in each redshift range in the other RA ranges are specified in Tables~\ref{dis_30_120} through~\ref{dis_30_300}. 

\begin{table}
{
\scriptsize
\begin{tabular}{|l|c|c|c|c|}
\hline
z &    cw    &  ccw  & ${cw}\over{cw+ccw}$ & P value \\      
\hline
0-0.05     &	245 &	 260 & 0.485	& 0.266  \\
0.05-0.1 &	442	& 429 & 0.507 &	0.34 \\
0.1-0.15 &	370 &	384 & 	0.491 &	0.32  \\
0.15-0.2 &	 229 &	185 & 	0.553 &	0.017  \\
0.2 - 0.5 &	 379 &	 302 &	 0.556 &	0.0018  \\
\hline
Total  & 1,665 & 1,560 & 0.516 & 0.034 \\
\hline
\end{tabular}
\caption{The distribution of the clockwise and counterclockwise SDSS galaxies in different redshift ranges in the RA range of $(30^o,120^o)$.}
\label{dis_30_120}
}
\end{table}

\begin{table}
{
\scriptsize
\begin{tabular}{|l|c|c|c|c|}
\hline
z &    cw    &  ccw  & ${cw}\over{cw+ccw}$ & P value \\      
\hline
0-0.05     &	1643 & 1635 & 0.501	& 0.451  \\
0.05-0.1 &	3351	& 3162 & 0.514 &	0.01 \\
0.1-0.15 &	2105 & 2115 & 0.499 &	0.445  \\
0.15-0.2 &	797 &	782 & 	0.500 &	0.362  \\
0.2 - 0.5 &	1165 & 1163 &  0.500 &	0.492  \\
\hline
Total  & 9,061 &  8,857 & 0.505 & 0.07 \\
\hline
\end{tabular}
\caption{The distribution of the clockwise and counterclockwise galaxies in different redshift ranges in the RA range of $(210^o,300^o)$.}
\label{dis_210_300}
}
\end{table}

\begin{table}
{
\scriptsize
\begin{tabular}{|l|c|c|c|c|}
\hline
z &    cw    &  ccw  & ${cw}\over{cw+ccw}$ & P value \\      
\hline
0-0.05     &	360 &	 402 & 0.472	& 0.068  \\
0.05-0.1 &	1040	& 1023 & 0.504 &	0.362 \\
0.1-0.15 &	714 &	735 & 	0.492 &	0.299  \\
0.15-0.2 &	322 &	379 & 	0.459 &	0.017  \\
0.2 -0.5 &	628 &	 680 &	 0.480 &	0.079  \\

\hline
Total  & 3,064 &  3,219 & 0.487 & 0.026 \\
\hline
\end{tabular}
\caption{The distribution of the clockwise and counterclockwise galaxies in different redshift ranges in the RA range of $(<30^o, >300^o)$.}
\label{dis_30_300}
}
\end{table}

The difference between the population of SDSS clockwise and counterclockwise galaxies in the different RA and redshift ranges is shown in Table~\ref{asymmetry_all}. The table shows the RA ranges in thinner slices of 30$^o$, and the redshift range slice is 0.15. The RA ranges of (60$^o$,90$^o$) and (270$^o$,300$^o$) are ignored due to the very low number of galaxies in them of 83 and 0, respectively. The error is determined by $\frac{1}{\sqrt{n}}$, such that n is the number of galaxies. Naturally, due to the separation of the data into many smaller sections, in most sections the number of galaxies does not allow statistical significance. However, the RA range of (120$^o$,180$^o$) shows statistically significant asymmetry between the population of clockwise and counterclockwise galaxies.

\begin{table*}
{
\tiny
\begin{tabular}{|l|c|c|c|c|c|c|c|c|c|c|}
\hline
RA &    0-30    &  30-60  & 90-120 & 120-150 & 150-180 & 180-210 & 210-240 & 240-270 & 300-330 & 330-360  \\      
\hline
$0<z<0.15$     &   -0.046$\pm$0.02	& -0.035$\pm$0.03   & 0.027$\pm$0.03  & 0.014$\pm$0.01  & -0.017$\pm$0.01  & 0.004$\pm$0.01  & 0.01$\pm$0.01  & 0.021$\pm$0.02 & 0.03$\pm$0.03 & 0.014$\pm$0.03    \\
$0.15<z<0.3$  &  -0.114$\pm$0.04 &	0.072$\pm$0.04 &	-0.031$\pm$0.07 &	-0.073$\pm$0.02 &	-0.053$\pm$0.02 &	0.03$\pm$0.02 &  0.035$\pm$0.02	& -0.045$\pm$0.04 & -0.086$\pm$0.07 & 0.007$\pm$0.04    \\
 $z>0.3$   &	  0.031$\pm$0.05   &	0.11$\pm$0.06   &		0.002$\pm$0.09   &	-0.131$\pm$0.03   &	-0.156$\pm$0.03   & 	-0.012$\pm$0.03   &	0.002$\pm$0.03   &	0.007$\pm$0.05   &	-0.052$\pm$0.13   &	-0.041$\pm$0.06       \\
Total     &	 -0.051$\pm$0.02	& 0.016$\pm$0.02	&	0.017$\pm$0.03 &	-0.015$\pm$0.01 &	-0.034$\pm$0.01 &	0.006$\pm$0.01 &	0.013$\pm$0.01 &	0.007$\pm$0.01	&	0.002$\pm$0.03 &	-0.003$\pm$0.02      \\
\hline
\# galaxies & 2,971 & 1,870 & 1,326 & 10,157 & 13,097 & 13,013 & 12,877 & 5,041 & 1,084 & 2,228 \\
\hline
\end{tabular}
\caption{Asymmetry (cw-ccw)/(cw+ccw) in different RA and redshift ranges.}
\label{asymmetry_all}
}
\end{table*}

\begin{figure}[h]
\includegraphics[scale=0.48]{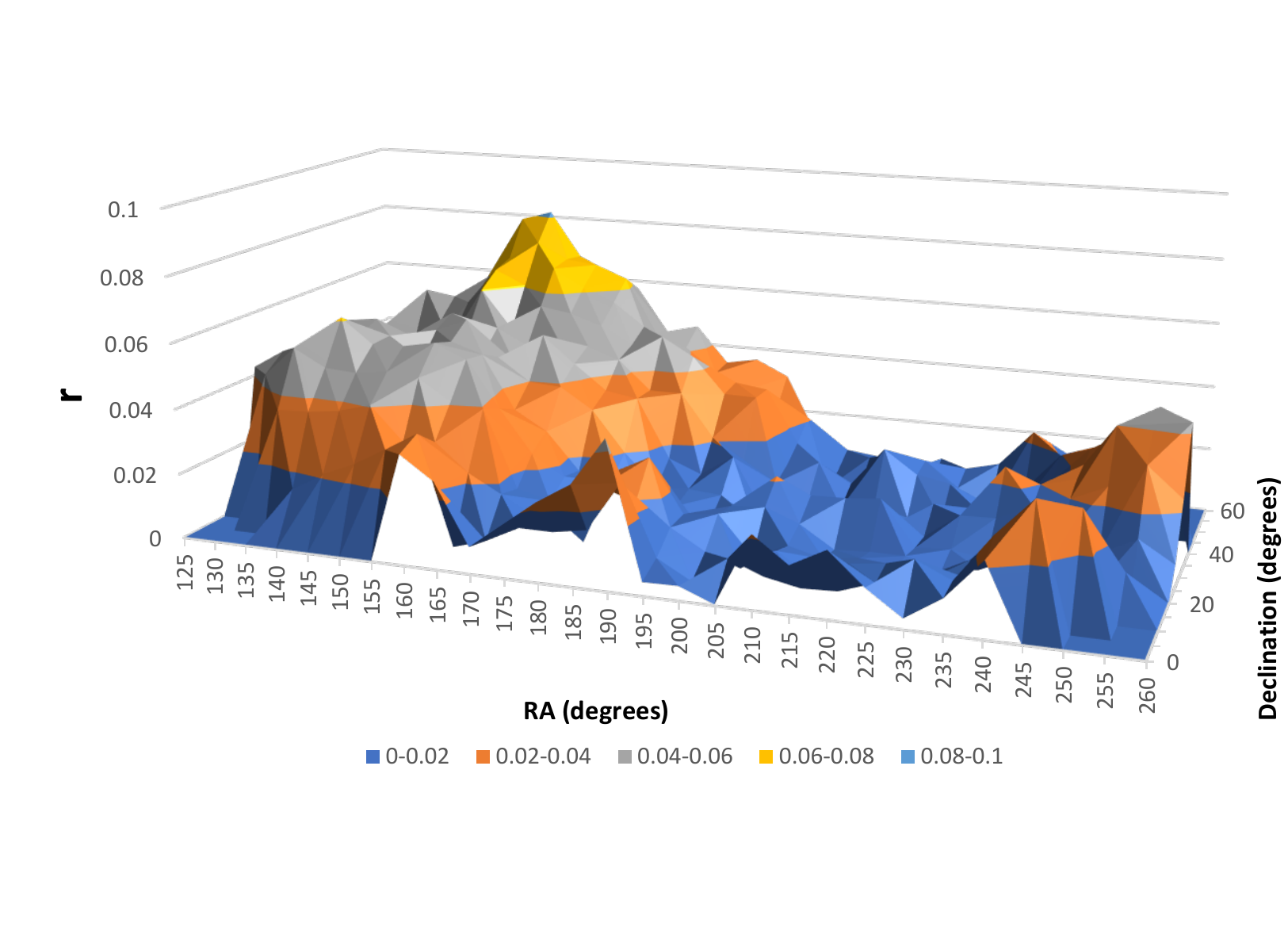}
\caption{The absolute value of the correlation between the redshift and the galaxy spin patterns in the 15$^o$ around different $(\alpha,\delta)$ combinations.}
\label{pearson_axis}
\end{figure}

To further analyze the correlation between the redshift and the distribution of galaxy spin directions, counterclockwise galaxies were assigned the value 1, and clockwise galaxies were assigned the value -1. Then, for every possible $(\alpha,\delta)$ combination (in increments of 5), the Pearson correlation between the galaxy spin pattern (1 or -1) and the redshift was computed for all galaxies that are 15$^o$ or less away from the $(\alpha,\delta)$ coordinates. In case the $(\alpha,\delta)$ coordinates had less than 3000 galaxies within 15$^o$ or less, the $(\alpha,\delta)$ coordinates were excluded. Figure~\ref{pearson_axis} shows the correlations in difference $(\alpha,\delta)$. The strongest correlation of 0.0815 (P$<$0.00001) was identified in (160,$^o$,50$^o$). The 1$\sigma$ range for the RA is (130$^o$,185$^o$), and for the declination it is (15$^o$,65$^o$). That, however, is limited to the sky covered by SDSS, as it is not possible to measure the correlation between the redshift and the rotation of direction of galaxies in unpopulated or underpopulated sky regions.  

The asymmetry between the number of clockwise and counterclockwise galaxies in SDSS changes with the direction of observation. A cosmological-scale dipole axis is expected to be observed in the form of cosine dependence with the direction of observation \citep{longo2011detection,shamir2012handedness,shamir2020large}. To identify the most likely dipole axis, for each $(\alpha,\delta)$ combination, the $\cos(\phi)$ of the galaxies was fitted into $d\cdot|\cos(\phi)|$, such that $\phi$ is the angular distance between the $(\alpha,\delta)$ coordinates and the coordinates of the galaxy, and d is the spin direction of the galaxy (1 for clockwise or -1 for counterclockwise) as was done in \citep{shamir2012handedness}. The probability of the axis was determined by assigning each galaxy with a random number within the set \{-1,1\}, and fitting $d|\cdot\cos(\phi)|$ to $\cos(\phi)$, such that $d$ is the randomly assigned spin direction (1 or -1). The $\chi^2$ was computed 2000 times for each $(\alpha,\delta)$ combination, and the mean and $\sigma$ were computed for each $(\alpha,\delta)$ combination. Then, the $\chi^2$ mean computed with the random spin directions was compared to the $\chi^2$ when $d$ was assigned to the actual spin direction of the galaxies. The difference (in terms of $\sigma$) between the $\chi^2$ of the actual spin patterns and the mean $\chi^2$ determined using randomly assigned spin patterns show the statistical likelihood of an axis at the $(\alpha,\delta)$ coordinates.

\begin{figure}[h]
\includegraphics[scale=0.14]{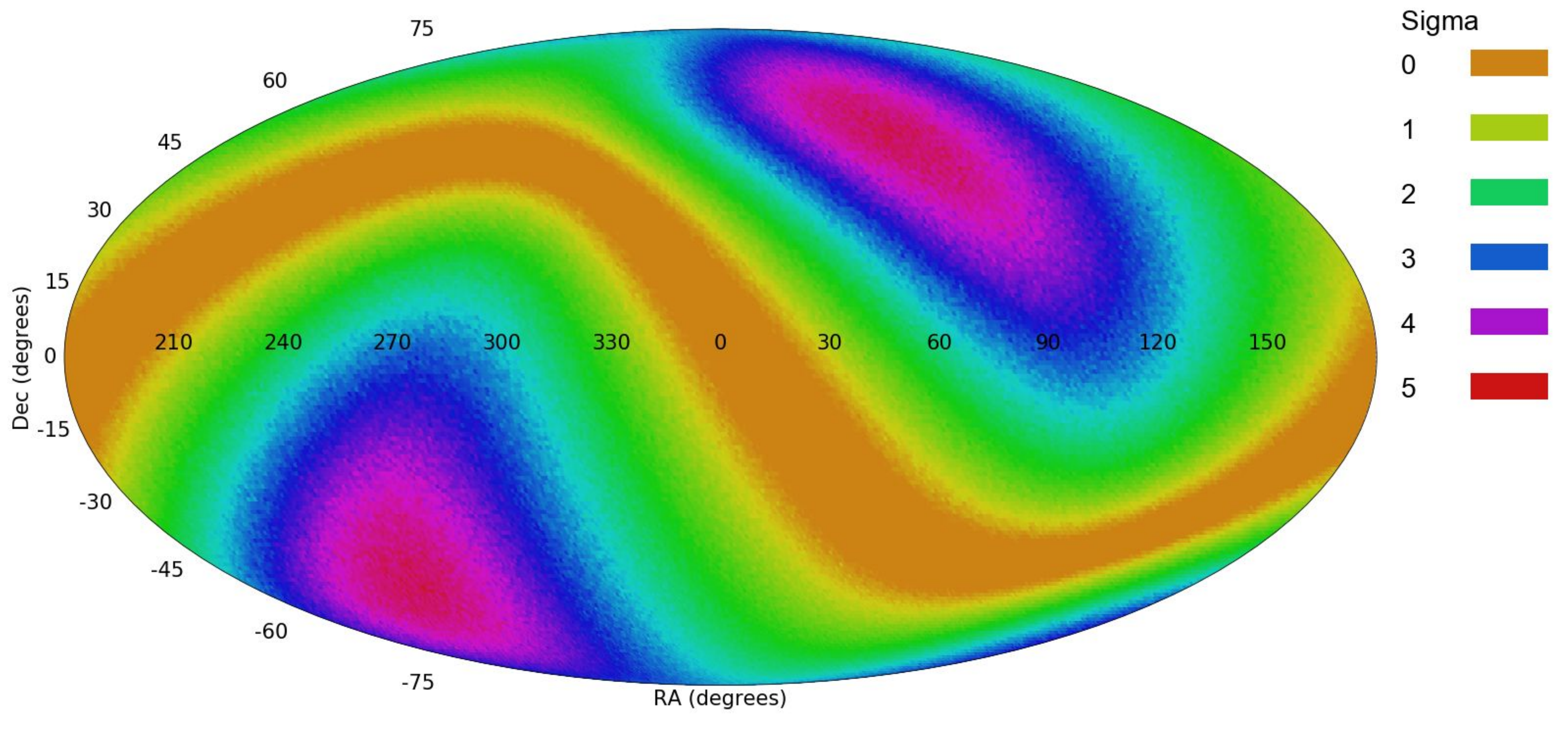}
\caption{The $\sigma$ of the likelihood of a dipole axis of SDSS galaxies in different $(\alpha,\delta)$ combinations.}
\label{axis}
\end{figure}

Figure~\ref{axis} shows the $\sigma$ for the asymmetry axis of all possible integer $(\alpha,\delta)$ combinations, and for different redshift ranges. The most likely $(\alpha,\delta)$ was identified at $(\alpha=69^o,\delta=56^o)$, with $\sigma$ of $\sim$4.63, meaning that the probability of such axis to occur by chance if the rotation directions of the SDSS galaxies are random is $(P<0.00001)$. The 1$\sigma$ error is $(19^o,107^o)$ for the right ascension, and $(25^o,77^o)$ for the declination. As Table~\ref{dis_120_210} shows, the asymmetry in SDSS galaxies changes across different redshift ranges. When identifying the most likely axis using just galaxies with $z>0.15$, the most likely axis is at $(\alpha=71^o,\delta=61^o)$, with a high $\sigma$ of $\sim$6.1. The 1$\sigma$ error is $(31^o,126^o)$ for the RA, and $(44^o,82^o)$ for the declination. However, when the galaxies are limited to $0<z<0.15$, the most likely axis is identified at $(53^o,37^o)$, with 1.9$\sigma$.

\begin{figure}[h]
\includegraphics[scale=0.28]{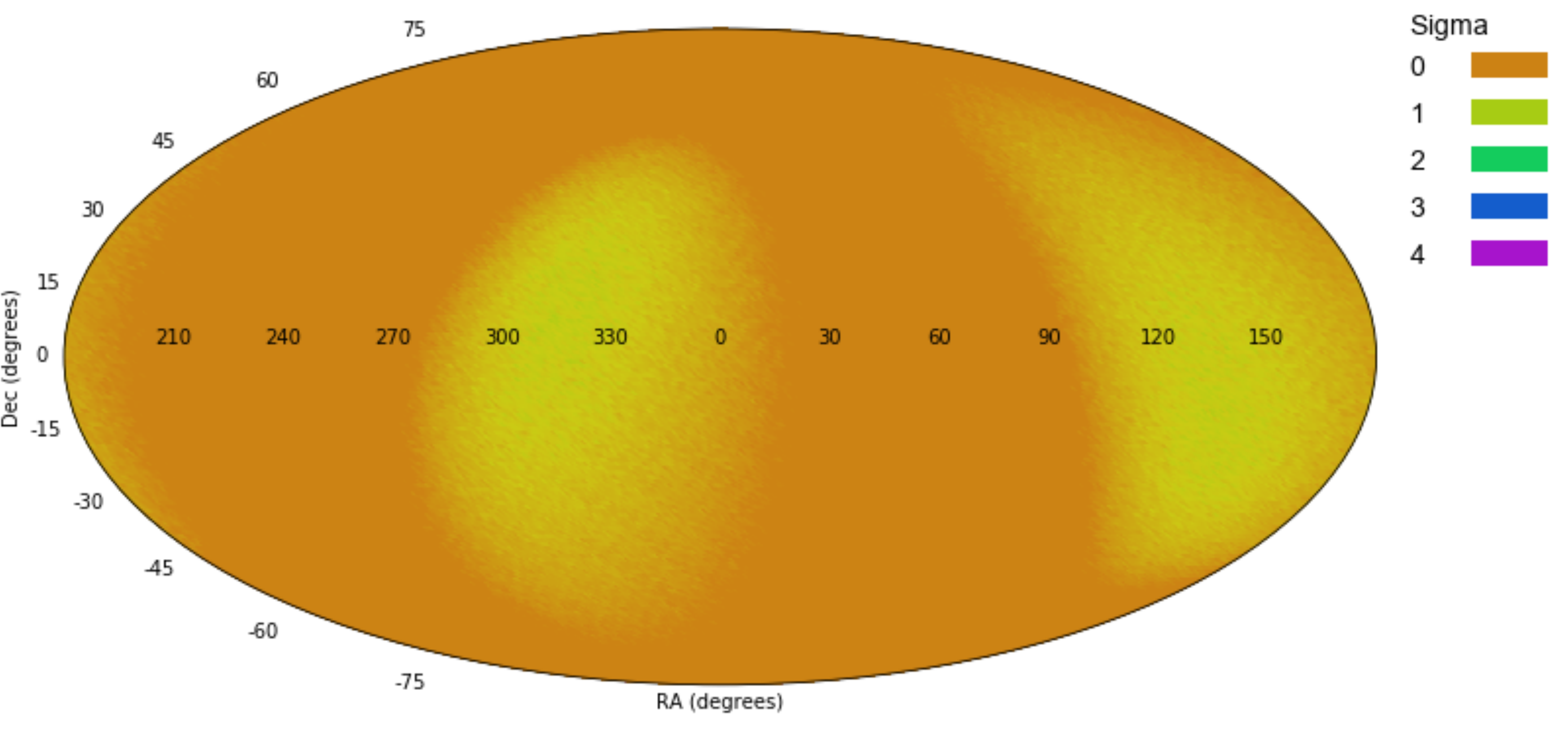}
\caption{The $\sigma$ of the axis of asymmetry in different $(\alpha,\delta)$ combinations such that galaxies were assigned with spin patterns randomly.}
\label{axis_random}
\end{figure}

Figure~\ref{axis_random} shows the result of the same experiment, but instead of using the spin patterns determined by Ganalyzer as described in Section~\ref{data}, each galaxy was assigned a random spin pattern (1 or -1). After assigning each galaxy with a random spin pattern, the attempt to identify the most likely axis was repeated. The probability of the most likely axis is just close to 1$\sigma$ ($\sim0.92\sigma$), and therefore does not show evidence of a dipole axis when the galaxies are assigned with random spin directions.

According to Table~\ref{directions}, the asymmetry shows more counterclockwise galaxies in RA ranges $(120^o,210^o)$ and $(>300^o, <30^o)$, and an excessive number of clockwise galaxies in the RA ranges of $(30^o,120^o)$ and $(210^o,300^o)$. That might be considered as certain evidence of a quadrupole alignment. Quadrupole alignment has also been noted in the context of CMB anisotropy as observed by COBE, WMAP, and Planck \citep{cline2003does,gordon2004low,zhe2015quadrupole}, suggesting the model of double inflation \citep{feng2003double,piao2004suppressing} and cosmological models that do not conform to the ``standard'' cosmology \citep{rodrigues2008anisotropic,piao2005possible,jimenez2007cosmology}. A more detailed discussion is provided in Section~\ref{conclusion}.

\begin{figure}[h]
\includegraphics[scale=0.14]{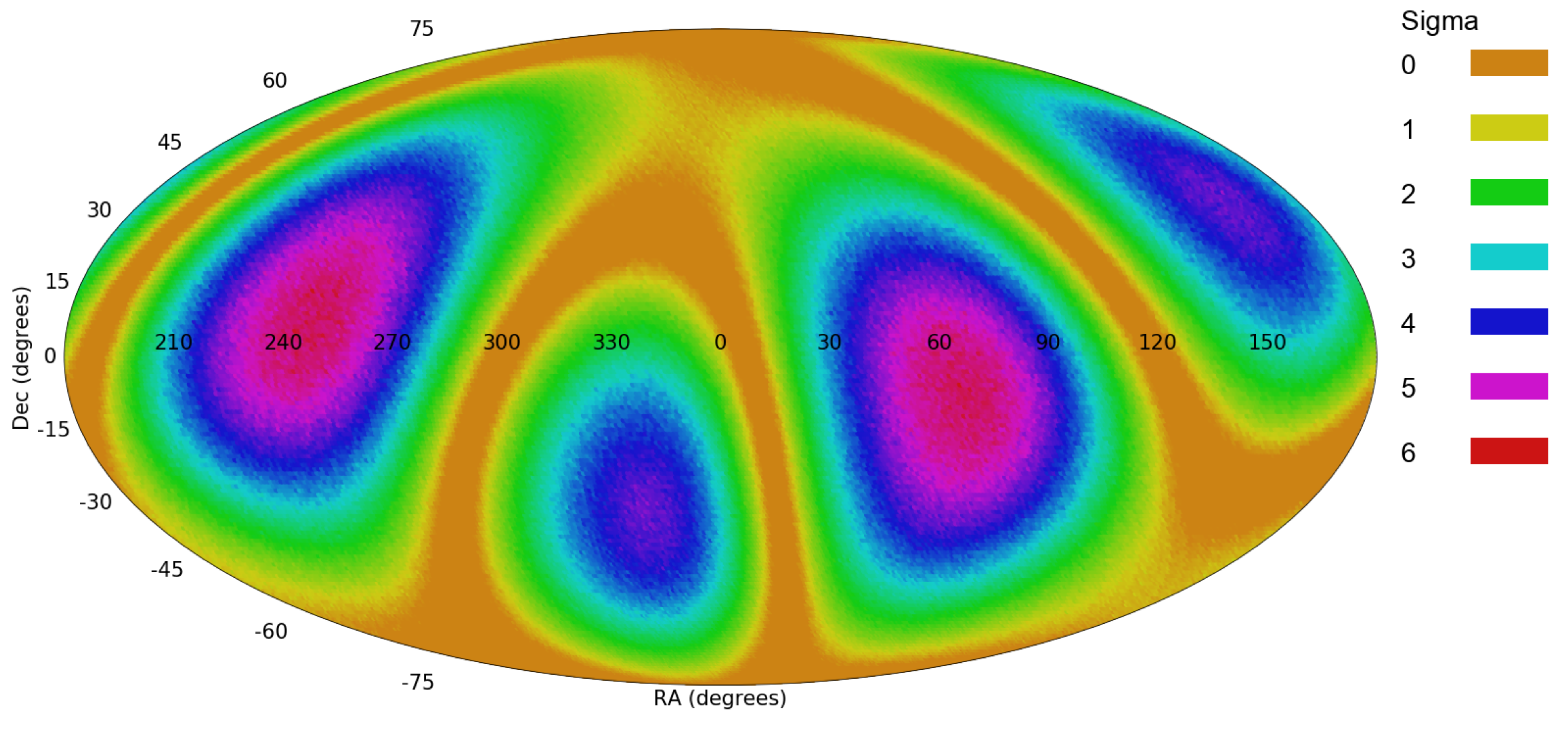}
\caption{The $\sigma$ of the likelihood of a quadrupole axis in different $(\alpha,\delta)$ combinations.}
\label{axis_quadrupole}
\end{figure}

Figure~\ref{axis_quadrupole} shows the $\chi^2$ fitting to cosine $2\phi$ dependence in each possible integer $(\alpha,\delta)$ combinations. The most likely axis is identified at $(\alpha=245^o,\delta=12^o)$, with certainty of $\sim5.13\sigma$. The 1$\sigma$ error range of the RA is $(225^o,271^o)$, and on the declination the error range is $(-11^o,31^o)$. Interestingly, that axis is within 1$\sigma$ error from the CMB cold spot \citep{bennett2003microwave} at $(48.77^o,-19.58^o)$. Another axis peaks at $(\alpha=158^o,\delta=43^o)$ with 3.97$\sigma$. That axis is close to the axis of highest correlation between redshift and the asymmetry between clockwise and counterclockwise galaxies as shown by Figure~\ref{pearson_axis}. Repeating the experiment by assigning the galaxies with random spin directions provided a profile similar to the profile displayed in Figure~\ref{axis_random}.

\begin{figure}[h]
\includegraphics[scale=0.14]{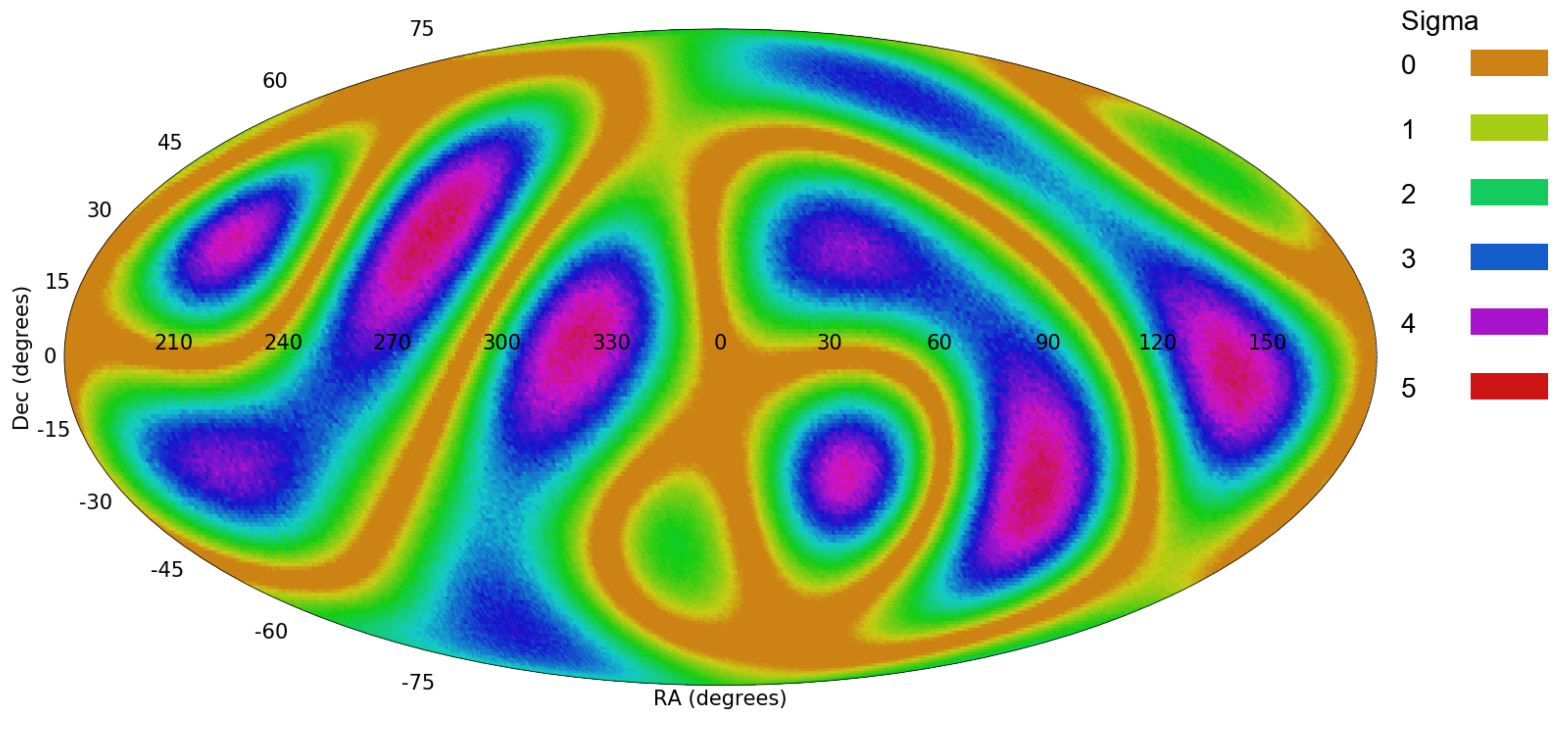}
\caption{The $\sigma$ of the likelihood of a octopole axis in different $(\alpha,\delta)$ combinations.}
\label{axis_octopole}
\end{figure}

An attempt to fit the spin directions of the galaxies to octopole alignment is displayed in Figure~\ref{axis_octopole}. The most likely axis is identified at $(\alpha=219^o,\delta=30^o)$ with probability of $\sigma\simeq4.86$. That probability is lower than the most likely axis when fitting to a quadrupole alignment.

\begin{figure}[h]
\includegraphics[scale=0.14]{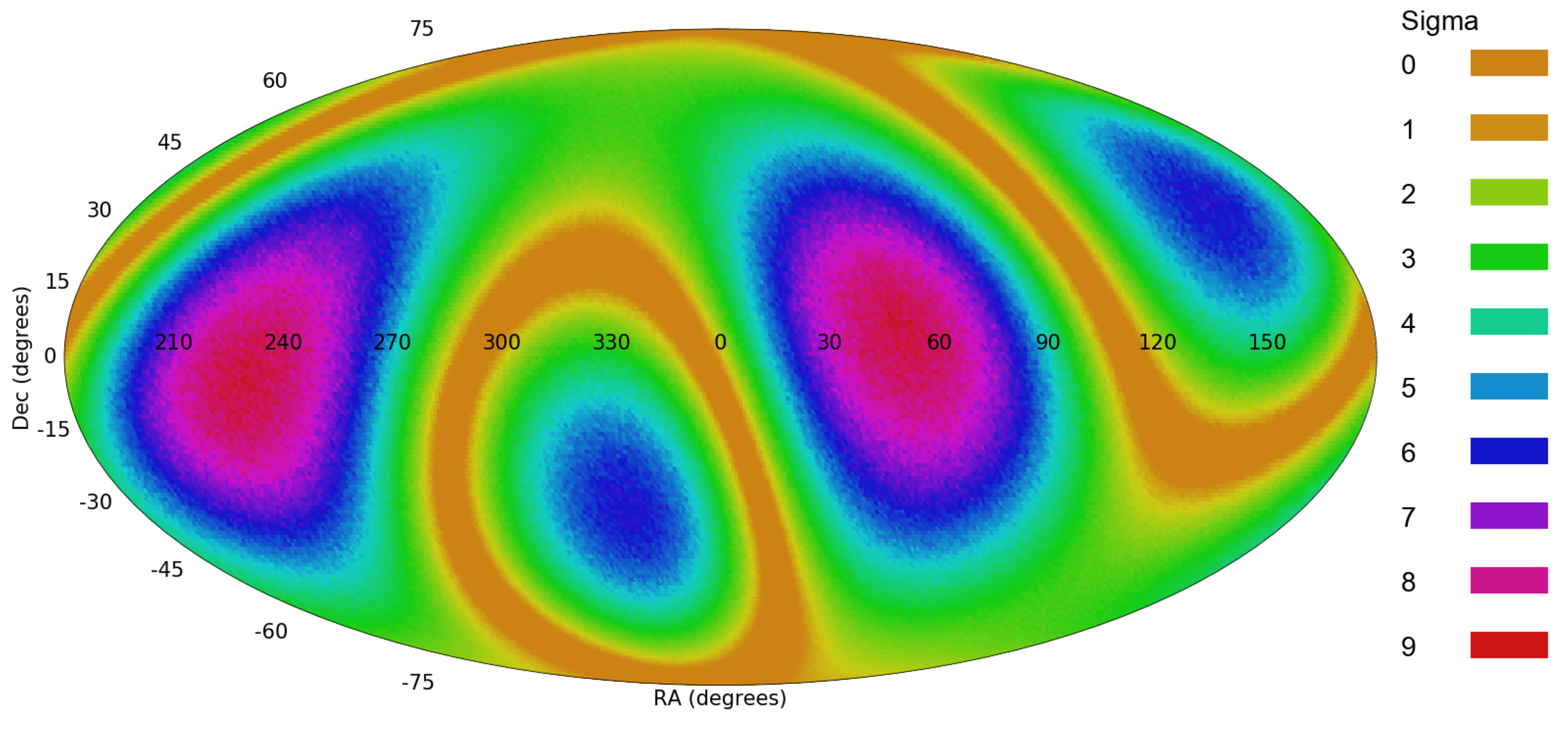}
\caption{The $\sigma$ of the likelihood of a quadrupole axis in different $(\alpha,\delta)$ combinations when all galaxies have $z>0.15$.}
\label{axis_quadrupole_z_015}
\end{figure}

Tables~\ref{dis_120_210} through~\ref{dis_30_300} show evidence that the asymmetry between galaxy spin patterns in SDSS increases with the redshift. When fitting the quadrupole using just galaxies with $(z>0.15)$, the most likely axes are at $(\alpha=232^o,\delta=7^o)$ and $(\alpha=153^o,\delta=34^o)$, roughly at the same location as when using the entire dataset, but the $\sigma$ of the most likely axes is much higher with 5.94 and 8.67, respectively. Figure~\ref{axis_quadrupole_z_015} shows the $\sigma$ in all possible $(\alpha,\delta)$ combinations. Although the most likely axes are roughly aligned with the most likely axes of the entire dataset of SDSS galaxies, the difference might suggest that it is possible that SDSS galaxies are aligned such that the asymmetry of their spin directions exhibits a certain drag.

\section{Comparison of the axes of asymmetry to Pan-STARRS data}
\label{panstarrs}

\begin{figure}[h]
\includegraphics[scale=0.60]{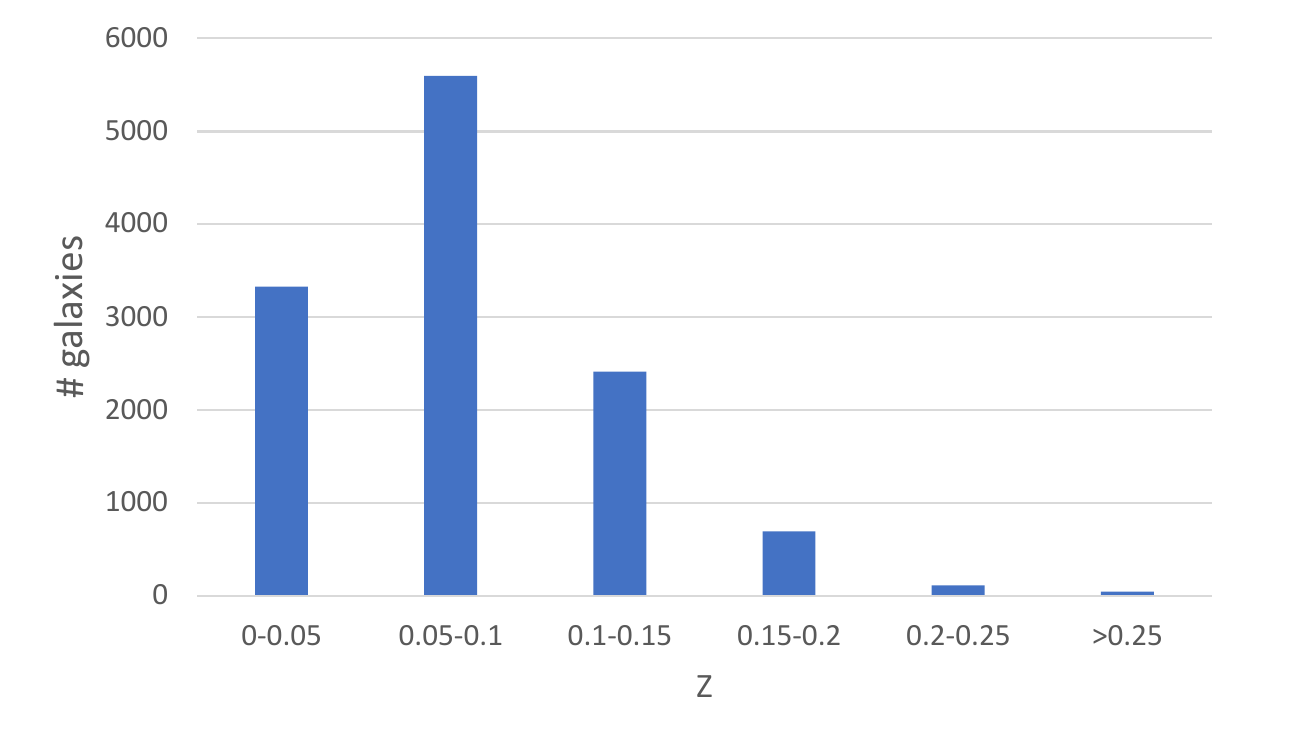}
\caption{The redshift distribution of the Pan-STARRS galaxies. Pan-STARRS is an optical sky survey that does not provide spectroscopy for the objects. Therefore, the redshift distribution was determined by using 12,186 objects that have corresponding spectroscopic objects in SDSS.}
\label{PanSTARRS_z_distribution}
\end{figure}

It is difficult to think of an error that can lead to the asymmetry described in Section~\ref{results}. The analysis process does not involve human intervention or machine learning that can lead to bias. Moreover, a software bias is expected to lead to consistent bias in all parts of the sky, rather than different bias in different parts of the sky, and different redshift ranges. It is also difficult to think of an error in SDSS that can lead to such asymmetry. But to test whether the asymmetry is in SDSS data, a comparison to Pan-STARRS data was made. For that purpose, a dataset of 2,394,452 Pan-STARRS objects identified as extended sources by all bands was used \citep{timmis2017catalog}. The galaxies were analyzed in the same manner described in Section~\ref{data}, providing a dataset of 33,028 galaxies classified by their spin patterns, collected in a process that is independent from the dataset of SDSS galaxies. The redshift distribution of the Pan-STARRS galaxies is shown in Figure~\ref{PanSTARRS_z_distribution}.


To compare the asymmetry identified in SDSS galaxies to Pan-STARRS galaxies, a subset of SDSS galaxies was selected such that the redshift distribution of the galaxies in that dataset matched the redshift distribution of the Pan-STARRS galaxies shown in Figure~\ref{PanSTARRS_z_distribution}. That led to a dataset of 38,998 SDSS galaxies with redshift distribution similar to the redshift distribution in Figure~\ref{PanSTARRS_z_distribution}. It is important to note that the SDSS galaxies were not selected so that the galaxies in both datasets are the same galaxies imaged by different telescopes, but were selected by their redshift so that the redshift distribution of both datasets was similar. While each of the two datasets was collected in an independent process, the footprints of SDSS and Pan-STARRS overlap, so some galaxies are expected to appear in both datasets. A comparison of the right ascension and declination of galaxies (to a distance of 0.001$^o$) in both datasets  showed that 4,426 galaxies are included in both the SDSS and Pan-STARRS datasets.

Table~\ref{DR14_PanSTARRS} shows the asymmetry in Pan-STARRS and SDSS in RA slices of 30$^o$. As expected, the asymmetry in the different RA slices is not identical due to statistical error, and possibly also due to differences in declination and different distribution of the galaxies inside these RA ranges. However, none of the RA ranges shows difference between the two datasets that is greater than 1$\sigma$, except for the RA range (0$^o$,30$^o$), where the difference between the two sky surveys is $\sim1.2\sigma$. When applying Bonferroni correction, no RA range has difference greater than 1$\sigma$. The sign of the asymmetry agrees in all RA ranges except for (240$^o$,270$^o$), which does not have a very high number of galaxies in both datasets. The Pearson correlation between the SDSS and Pan-STARRS asymmetries in the different RA slices is $\sim$0.55, and the mere chance probability for such correlation is P$\simeq$0.05. The two most populated sections in both datasets, (150$^o$,180$^o$) and (180$^o$,210$^o$), have near identical asymmetries.

\begin{table}[ht]
{
\scriptsize
\begin{tabular}{|l|c|c|c|c|}
\hline
RA &   ${cw-ccw}\over{cw+ccw}$ & \# galaxies &  ${cw-ccw}\over{cw+ccw}$   &  \# galaxies  \\ 
     &  SDSS                                 &     SDSS     &   PS                       &   PS  \\
\hline
0$^o$-30$^o$          &  -0.049 &  2596 &   -0.009  &  3559  \\
30$^o$-60$^o$        &  -0.001 & 1497  &   -0.031  &  2676  \\
60$^o$-90$^o$        &  -0.115 &  61 &   -0.015  &   1698 \\
90$^o$-120$^o$       &  0.024 &  1152 &   0.041  &   1099 \\
120$^o$-150$^o$     &  0.001 &  8550 &  0.025   &   3473 \\
150$^o$-180$^o$     &  -0.020 &  10482 &  -0.022   &   5064 \\
180$^o$-210$^o$     &  -0.001 &  7676 &   -0.001  &   5195 \\
210$^o$-240$^o$     &  0.014 &  5892 &   0.023  &   4088 \\
240$^o$-270$^o$     &  -0.004 &  1018 &   0.013  &   1874 \\
270$^o$-300$^o$     &  -        &  0       &   0.082  &   429 \\
300$^o$-330$^o$     &  -0.081 & 74 &   -0.024  &   1074 \\
330$^o$-360$^o$     &  - &  0 &   -0.007  &   2799 \\
\hline
\end{tabular}
\caption{The asymmetry between clockwise and counterclockwise galaxies in different RA ranges in Pan-STARRS and SDSS. The SDSS galaxies are selected such that their redshift distribution matches the redshift distribution of the Pan-STARRS galaxies. For all RA ranges the difference is within 1$\sigma$, except for (0$^o$,30$^o$), where the difference is $\sim1.2\sigma$. The probability to have that correlation by chance is P$\simeq$0.05.}
\label{DR14_PanSTARRS}
}
\end{table}

\begin{figure}[h]
\includegraphics[scale=0.14]{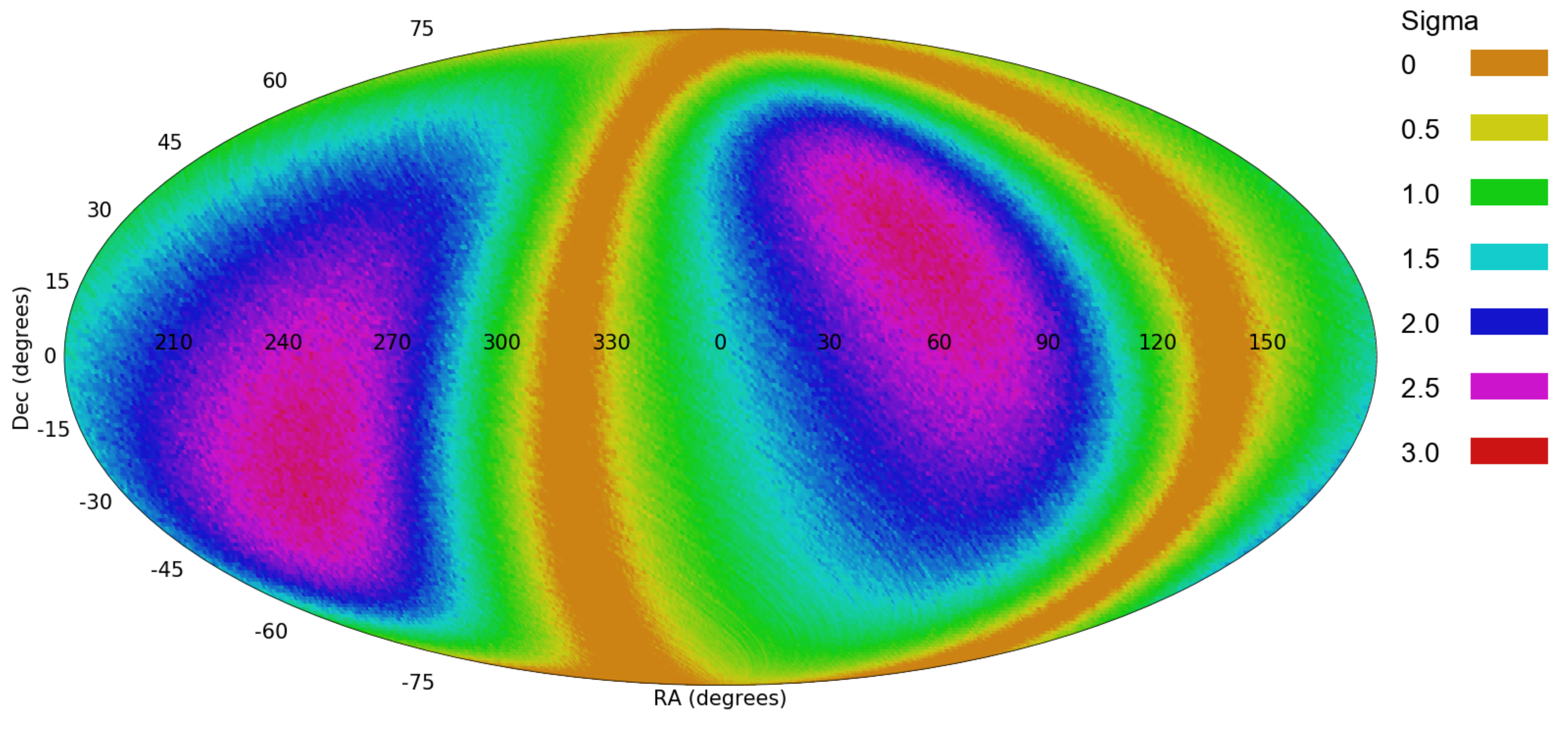}
\caption{The probability of asymmetry dipole axis in all possible ($\alpha$,$\delta$) combinations in SDSS data such that the redshift distribution is similar to the redshift distribution of the Pan-STARRS galaxies.}
\label{DR14_PanSTARRS_dipole}
\end{figure}

\begin{figure}[h]
\includegraphics[scale=0.14]{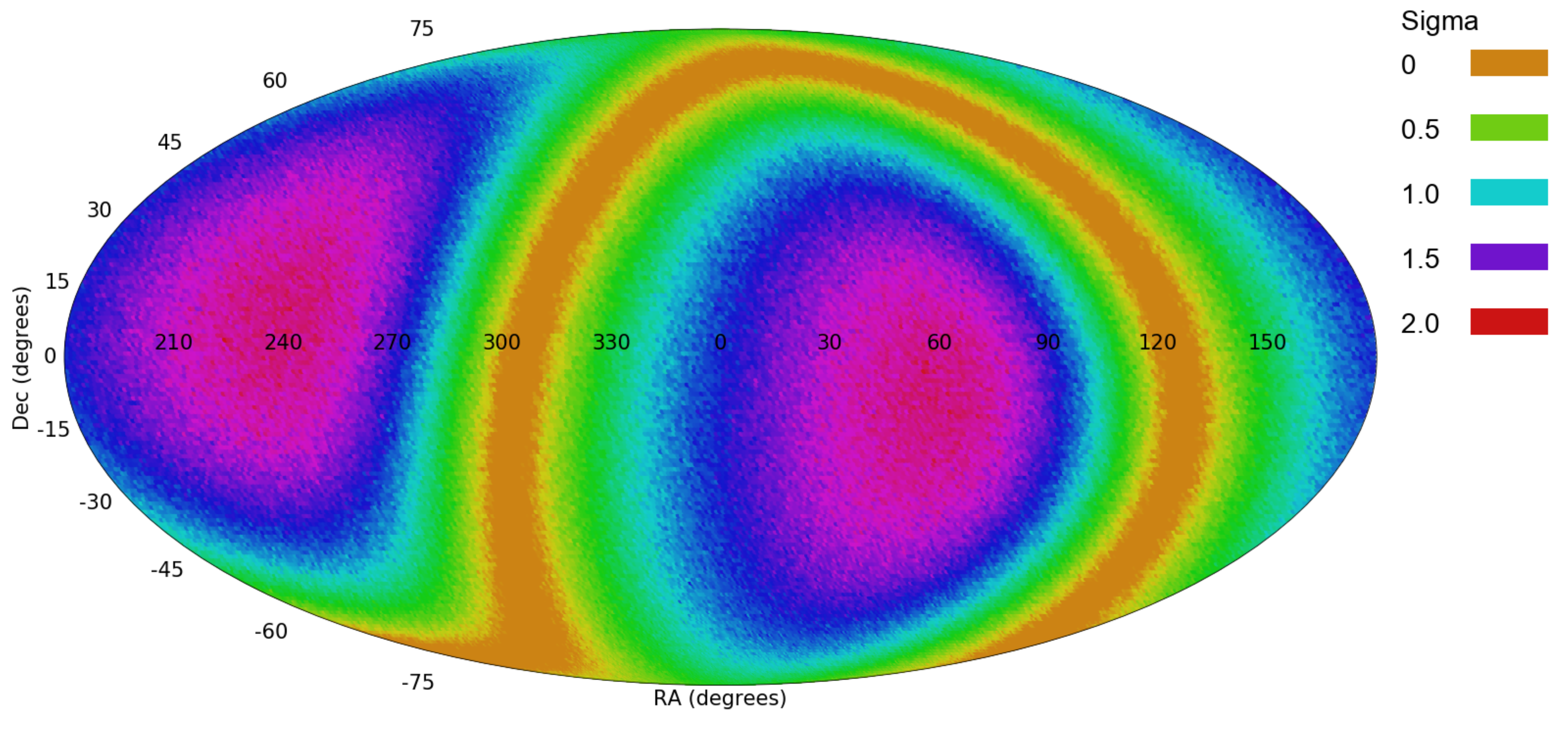}
\caption{The probability of asymmetry dipole axis in all possible ($\alpha$,$\delta$) combinations in Pan-STARRS galaxies.}
\label{PanSTARRS_DR14_dipole}
\end{figure}

To compare the most likely dipole axis of asymmetry in Pan-STARRS galaxies to the most likely axis of asymmetry in SDSS galaxies, the two datasets used in Table~\ref{DR14_PanSTARRS} was used in the same manner described in Section~\ref{results}. Figures~\ref{DR14_PanSTARRS_dipole} and~\ref{PanSTARRS_DR14_dipole} show the probability of the most likely axis of asymmetry in each ($\alpha$,$\delta$) combination, in SDSS and Pan-STARRS, respectively. The most likely axis in SDSS is at $(\alpha=49^o,\delta=21^o)$, with probability of $\sim2.05\sigma$. The right ascension is nearly identical to the most likely axis identified in the Pan-STARRS data at $\alpha=47^o$, with $\sigma\simeq1.87$ (one-tail P$\simeq$0.03). The declination shows a certain difference of 21$^o$ in SDSS compared to -1$^o$ in Pan-STARRS. However, the declination range of both SDSS and Pan-STARRS is much more narrow compared to the range of the RA, and that can lead to higher error in the estimation of the declination of the most likely axis. In any case, the $1\sigma$ error range of the declination in Pan-STARRS is $(-78^o<\delta<43^o)$, so the difference between the two datasets is well within statistical error. The difference between declination of -1 and declination of 21 in the Pan-STARRS data is merely $\sim0.15\sigma$.

\begin{figure}[h]
\includegraphics[scale=0.48]{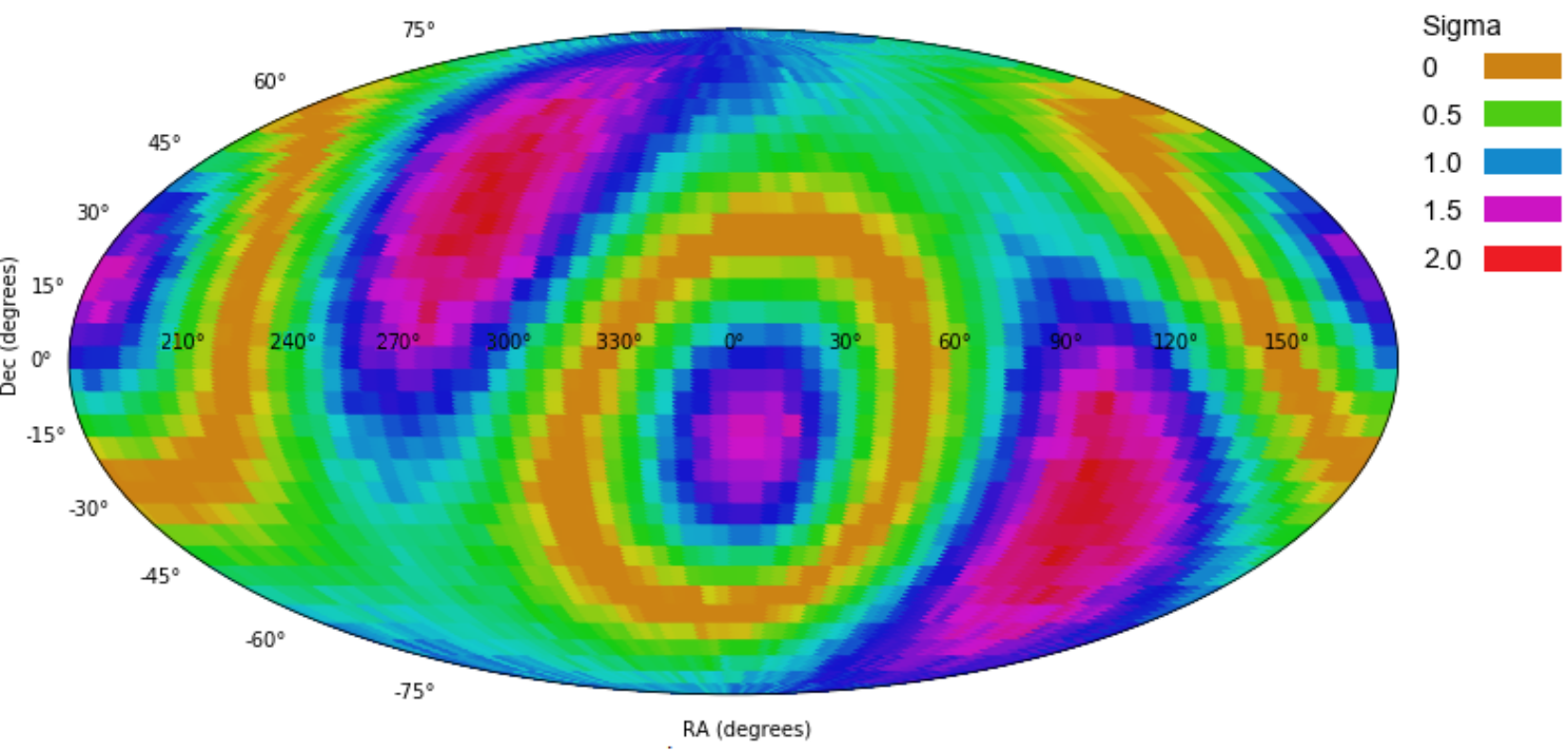}
\caption{The probability quadrupole alignment in all possible ($\alpha$,$\delta$) combinations in SDSS data such that the redshift distribution is similar to the redshift distribution of the Pan-STARRS galaxies.}
\label{DR14_PanSTARRS_quad}
\end{figure}

\begin{figure}[h]
\includegraphics[scale=0.14]{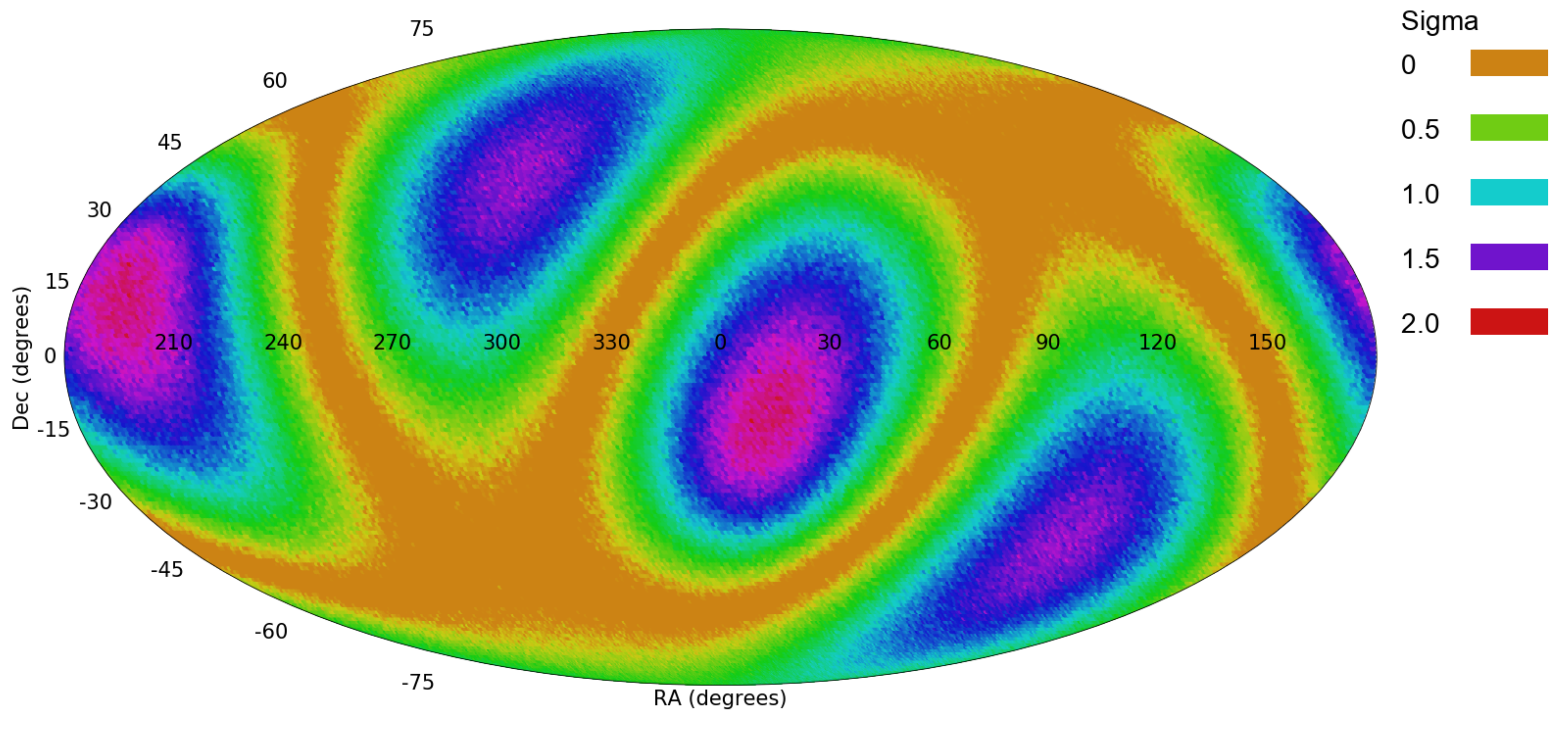}
\caption{The probability of quadrupole alignment in all possible ($\alpha$,$\delta$) combinations in Pan-STARRS galaxies.}
\label{PanSTARRS_DR14_quad}
\end{figure}

The Pan-STARRS data was also used to compare the quadrupole alignments in both datasets. Figures~\ref{DR14_PanSTARRS_quad} and~\ref{PanSTARRS_DR14_quad} show the most likely quadrupole alignment in each ($\alpha$,$\delta$) in SDSS and Pan-STARRS, respectively. The RA of the most likely axis in SDSS is at $(\alpha=7^o,\delta=4^o)$, with $\sigma\simeq1.81$. That axis is well within the $1\sigma$ error range of the most likely axis shown in the Pan-STARRS data, peaking at $(\alpha=17^o,\delta=-2^o)$ with $\sigma$ of 1.64 (one-tail P$\simeq$0.05), and the 1$\sigma$ error range is $(339^o<\alpha<85^o)$ for the right ascension and $(-42^o,36^o)$ for the declination.

\section{Conclusion}
\label{conclusion}

Previous studies showed evidence of differences between galaxies with clockwise and counterclockwise spin patterns \citep{shamir2013color,hoehn2014characteristics,shamir2016asymmetry,shamir2017colour,shamir2017photometric,shamir2017large}. While some previous attempts were limited by the bias of the human eye \citep{land2008galaxy}, model-driven automatic morphological analysis of galaxies was able to process much larger datasets that are not biased by the human perception, showing evidence that the distribution of clockwise and counterclockwise galaxies as seen from Earth is not uniform in all parts of the sky, and correspond to the direction of observation \citep{shamir2012handedness}.

Here a dataset of $\sim6.4\cdot10^4$ galaxies with spectra is used to show an asymmetry in the distribution of clockwise and counterclockwise galaxies as observed by SDSS. The entire process of data analysis is machine-based, and therefore no human bias can affect the results. Moreover, all algorithms are model-driven and are based on defined rules, and none of the stages of data annotation involves machine learning. In particular, the analysis process does not involve deep neural networks, which use highly complex rules and can be considered ``black box''. These data analysis systems can capture instrumental bias that is difficult to identify due to the highly non-trivial nature of the rules. In this study no human bias or machine learning was used, and therefore no human bias or other effects can lead to bias in the training set, as no training data is used in any stage of the data analysis process. The data annotation is deterministic, and does not involve complex non-intuitive set of automatically-generated rules that are often used by machine learning systems. Analysis of Pan-STARRS galaxy data shows that when the redshift distribution of SDSS galaxies is similar to the redshift distribution of Pan-STARRS galaxies, Pan-STARRS galaxies show very similar positions of the most likely dipole and quadrupole axes, well within statistical error.

As discussed above, it is difficult to think of an atmospheric or instrumental effect that would exhibit itself in the form presented in this paper. However, digital sky surveys are some of the more complex research instruments in existence. These systems combine advanced autonomous instruments with sophisticated data analysis pipelines, making it difficult to predict all aspects of the final data products that they generate. It is therefore a possibility that the observed asymmetry is related to an unknown flaw in the instrument or processing pipeline rather than an astronomical effect.

The analysis of the data shows that the number of clockwise galaxies imaged by SDSS is different from the number counterclockwise galaxies observed from Earth by SDSS, and the difference in the number of clockwise and counterclockwise galaxies changes based on the direction of observation. The difference also changes with the redshift, providing evidence that the asymmetry between the number of clockwise and counterclockwise galaxies increases as the redshift gets higher. The distribution of the spin directions of galaxies imaged by SDSS  forms an axis with probability far higher than mere chance, and gets to over 8$\sigma$. The distribution of the spin pattern best fits a quadrupole alignment, and the most likely axis is interestingly close to the CMB cold spot. The spin pattern of a spiral galaxy is also an indication of its spin direction \citep{iye2019spin}, and therefore the results show that in the redshift range that can be observed by SDSS (with identifiable galaxy morphology) the earlier universe is more homogeneous in terms of the distribution of galaxy spin directions.

It is naturally difficult to explain the observation using the standard models. Existing literature suggests that the change of the spin direction asymmetry with the redshift can be related to subtle primordial chiral violation carried to the current epoch through spin direction asymmetry \citep{yu2020probing}. Another explanation is related to parity-breaking gravitational waves, which can affect galaxy shape during inflation \citep{biagetti2020primordial}. These theories provide explanations that agree with the observation described in this paper, yet without necessarily violating the foundational cosmological assumptions. Further studies using higher redshifts will be needed to determine whether the anisoptropy is of primordial nature.

Fitting the distribution of the spin directions of galaxies imaged by SDSS and Pan-STARRS reveals multipole alignment. Anisotropy has been observed with cosmic radiation \citep{smoot1977detection}, and large-scale quadrupole alignment has been observed by CMB data \citep{cline2003does,gordon2004low,zhe2015quadrupole}, leading to cosmological theories that shift from the standard cosmological models \citep{feng2003double,piao2004suppressing,rodrigues2008anisotropic,piao2005possible,jimenez2007cosmology,bohmer2008cmb}. It also led to the ellipsoidal universe model \citep{campanelli2006ellipsoidal,campanelli2007cosmic,gruppuso2007complete}, and geometrical models of the universe that can explain the quadrupole alignment \citep{weeks2004well,efstathiou2003low}. As the spin pattern of a spiral galaxy is also an indication of its spin direction \citep{iye2019spin}, the distribution of the spin directions in such a large scale can be an indication of a rotating universe \citep{godel1949example,ozsvath1962finite,ozsvath2001approaches,sivaram2012primordial,chechin2016rotation}.

The spin pattern of a galaxy is a crude measurement, and there is no known atmospheric or other effect that can make a clockwise galaxy seem counterclockwise or vice versa. The classification of the galaxies was done in a fully automatic manner, and no human bias can have any impact on the results. The algorithm used for the classification is a model-driven algorithm that uses clear rules for the annotations. It is not driven by a machine learning process, in which the training data can add bias to the result, and the complex nature of the rules that determine the output makes it difficult to assess the way the classification is made. The asymmetry in SDSS and Pan-STARRS galaxies is identified in different parts of the sky that are in opposite hemispheres. Previous work showed with statistical significance that the asymmetry between clockwise and counterclockwise galaxies changes with the direction of observation \citep{shamir2017colour,shamir2017photometric,shamir2017large}. These results had very good agreement across two different sky surveys: SDSS and PanSTARRS \citep{shamir2017large}. That observation further eliminates the possibility of a software error, as such error should have exhibited itself in the form of consistent asymmetry in all parts of the sky, regardless of the direction of observation. As Table~\ref{directions} shows, the direction of asymmetry is different in different parts of the sky, such that some parts show a higher number of clockwise galaxies, and other parts of the sky show a higher number of counterclockwise galaxies. A computer error would have expected to exhibit itself in the form of a consistent bias, and would not invert in different parts of the sky, that are also in different hemispheres. Tables~\ref{dis_120_210} through~\ref{dis_30_300} show that the asymmetry of galaxies imaged by SDSS increases with the redshift, and the asymmetry grows in opposite directions in different parts of the sky. That also indicates that the asymmetry is not the result of a software error, as a software error would be consistent in all parts of the sky, and would lead to a decreasing asymmetry with the redshift due to the increasing difficulty in classifying high redshift galaxies that tend to be smaller and dimmer than galaxies with low redshift. Milky Way obscuration is also not expected to cause asymmetry between clockwise and counterclockwise galaxies, as such obscuration is expected to have the same effect on both types galaxies. It is therefore unlikely that the obscuration of the Milky Way would lead to an excessive number of either clockwise or counterclockwise galaxies, as both should be obscured in an equal manner.

As Figure~\ref{distribution} shows, the span of the galaxies used in this study is far larger than any known astrophysical structure, and might therefore provide evidence for violation of the isotropy assumption. While the isotropy and homogeneity assumptions are pivotal to standard cosmological models, these assumptions have not been proven, and spatial homogeneity cannot be verified directly \citep{ellis1979homogeneity}. Some evidence of isotropy violation at the cosmological scale have been observed through other messengers such as gamma ray bursts \citep{meszaros2019oppositeness}, radio sources \citep{bengaly2018probing}, Ia supernova \citep{javanmardi2015probing}, distribution of galaxy morphology \citep{javanmardi2017anisotropy}, and cosmic microwave background \citep{aghanim2014planck,hu1997cmb,cooray2003cosmic,ben2012parity,eriksen2004asymmetries}. The distribution of spin directions shown here is at the scale far larger than any known astrophysical structure, and challenges the standard cosmological models \citep{kroupa2012dark}. Future and more powerful sky surveys such as the Vera Rubin Telescope, or combination of optical surveys such as the DECam Legacy Survey \citep{blum2016decam} with powerful spectroscopy surveys such as DESI \citep{font2014desi} can provide a higher resolution profiling of the asymmetry.


\section*{Acknowledgments}

This study was supported in part by NSF grants AST-1903823 and IIS-1546079. Funding for the Sloan Digital Sky Survey IV has been provided by the Alfred P. Sloan Foundation, the U.S. Department of Energy Office of Science, and the Participating Institutions. SDSS-IV acknowledges support and resources from the Center for High-Performance Computing at the University of Utah. The SDSS web site is www.sdss.org.

SDSS-IV is managed by the Astrophysical Research Consortium for the Participating Institutions of the SDSS Collaboration including the Brazilian Participation Group, the Carnegie Institution for Science, Carnegie Mellon University, the Chilean Participation Group, the French Participation Group, Harvard-Smithsonian Center for Astrophysics, Instituto de Astrof\'isica de Canarias, The Johns Hopkins University, Kavli Institute for the Physics and Mathematics of the Universe (IPMU) / 
University of Tokyo, the Korean Participation Group, Lawrence Berkeley National Laboratory, Leibniz Institut f\"ur Astrophysik Potsdam (AIP), Max-Planck-Institut f\"ur Astronomie (MPIA Heidelberg), Max-Planck-Institut f\"ur Astrophysik (MPA Garching), Max-Planck-Institut f\"ur Extraterrestrische Physik (MPE), National Astronomical Observatories of China, New Mexico State University, New York University, University of Notre Dame, Observat\'ario Nacional / MCTI, The Ohio State University, Pennsylvania State University, Shanghai Astronomical Observatory, United Kingdom Participation Group, Universidad Nacional Aut\'onoma de M\'exico, University of Arizona, University of Colorado Boulder, University of Oxford, University of Portsmouth, University of Utah, University of Virginia, University of Washington, University of Wisconsin, Vanderbilt University, and Yale University.

The Pan-STARRS1 Surveys (PS1) and the PS1 public science archive have been made possible through contributions by the Institute for Astronomy, the University of Hawaii, the Pan-STARRS Project Office, the Max-Planck Society and its participating institutes, the Max Planck Institute for Astronomy, Heidelberg and the Max Planck Institute for Extraterrestrial Physics, Garching, The Johns Hopkins University, Durham University, the University of Edinburgh, the Queen's University Belfast, the Harvard-Smithsonian Center for Astrophysics, the Las Cumbres Observatory Global Telescope Network Incorporated, the National Central University of Taiwan, the Space Telescope Science Institute, the National Aeronautics and Space Administration under Grant No. NNX08AR22G issued through the Planetary Science Division of the NASA Science Mission Directorate, the National Science Foundation Grant No. AST-1238877, the University of Maryland, Eotvos Lorand University (ELTE), the Los Alamos National Laboratory, and the Gordon and Betty Moore Foundation.

\bibliographystyle{apalike}


\bibliography{Liorshamir_ms}

\end{document}